# Second-harmonic generation and the conservation of spatiotemporal orbital angular momentum of light

Guan Gui,[1] Nathan J. Brooks,[1] Henry C. Kapteyn,[1,2] Margaret M. Murnane,[1] and Chen-Ting Liao[1*]

[1]JILA, University of Colorado and NIST, 440 UCB, Boulder, Colorado 80309, USA
[2]KMLabs Inc., 4775 Walnut Street, Suite 102, Boulder, Colorado 80301, USA
*chenting.liao@colorado.edu

## Abstract

Light with spatiotemporal orbital angular momentum (ST-OAM) is a recently discovered type of structured and localized electromagnetic field. This field carries characteristic space-time spiral phase structure and transverse intrinsic OAM. In this work, we present the generation and characterization of the second-harmonic of ST-OAM pulses. We uncovered the conservation of transverse OAM in a second-harmonic generation process, where the space-time topological charge of the fundamental field is doubled along with the optical frequency. Our experiment thus suggests a general ST-OAM nonlinear scaling rule— analogous to that in conventional OAM of light. Furthermore, we observe that the topology of a second-harmonic ST-OAM pulse can be modified by complex spatiotemporal astigmatism, giving rise to multiple phase singularities separated in space and time. Our study opens a new route for nonlinear conversion and scaling of light carrying ST-OAM with the potential for driving other secondary ST-OAM sources of electromagnetic fields and beyond.

## Introduction

The orbital angular momentum (OAM) of light is a type of angular momentum associated with wavefront or phase vortices in the electromagnetic field[1,2]. For a propagating paraxial wave, the longitudinal OAM of light means that the OAM is parallel to the averaged wavevector and propagation direction of the beam. OAM can be intrinsic or extrinsic (c.f., a recent review in Ref[3])– intrinsic OAM implies that the angular momentum is reference frame independent and can be described by an integer quantum number called (spatial) topological charge $\ell_s$. The intrinsic longitudinal OAM (here referred to as conventional OAM) of light has an OAM of $\hbar\ell_s$ per photon and a spiral phase $e^{i\ell_s\Phi_s(x,y)}$ surrounding a phase singularity in the $(x,y)$ plane (Fig. 1a). Most studies over the past three decades have been focusing on conventional OAM, which has impacted many important applications including optical tweezers, super-resolution imaging, quantum and classical communication, scatterometry-based surface metrology[4–6]. Very recently, time-varying OAM of light has been discovered[7].

In contrast, a transverse OAM of light implies that the OAM is perpendicular to the averaged wavevector of the beam. This means that a spiral phase resides in space-time, e.g., the $(x,t)$ plane (or equivalently, the $(x,z)$ plane) in a simplified 2D case (Fig. 1a), and is thus referred to as spatiotemporal orbital angular momentum (ST-OAM). By analogy with conventional OAM, we can designate an integer $\ell$ as the spatiotemporal topological charge to describe the space-time winding phase $e^{i\ell\Phi(x,z,t)}$. The scalar field carrying ST-OAM reads

$$E(x,y,z,t) \propto E_0(x,y,z)e^{i\ell\Phi(x,z,t)}e^{i(k_z z-\omega t)}, \qquad (1)$$

where $E_0$ is the scalar field envelope, $\omega$ is the optical frequency, and $\Phi$ is the spatiotemporal phase. This transverse OAM of light was theoretically predicted[3,8,9], and observed in filamentation[10]. Very recently, optical pulses with ST-OAM were experimentally realized in the linear regime for the first time[11,12]. Because that ST-OAM of light was only recently discovered with few experimental observations, many of its properties are still elusive. Therefore, the extent to which they can be described by an analogy with the conventional OAM of light remains unclear.

In nonlinear frequency conversion, the conventional OAM of light follows a simple scaling rule, where the $N$-th harmonic has a topological charge $N\ell_s$, reflecting OAM conservation. This rule has been verified for second-harmonic generation (SHG)[13,14], non-perturbative high-order ($N > 10$) harmonic generation[15], and can be generalized to describe sum- and difference-frequency generation processes[16]. This rule only applies to scalar fields without spin angular momentum, otherwise, total angular momentum conservation must be included[17].

Here we experimentally investigate the behavior of ST-OAM pulses during the frequency up-conversion process of SHG. By uncovering the conservation of ST-OAM in an SHG process from $\ell = 1$ to $\ell = 2$, where the space-time topological charge of the fundamental field is doubled along with the optical frequency, our experiment demonstrates a general ST-OAM nonlinear scaling rule—analogous to that describing the (spatial) topological charges in conventional OAM of light. We also investigate the effects of spatiotemporal astigmatism in SHG, which leads to non-conserved topological changes of the spiral phase structure, and the creation of multiple phase singularities separated in space-time.

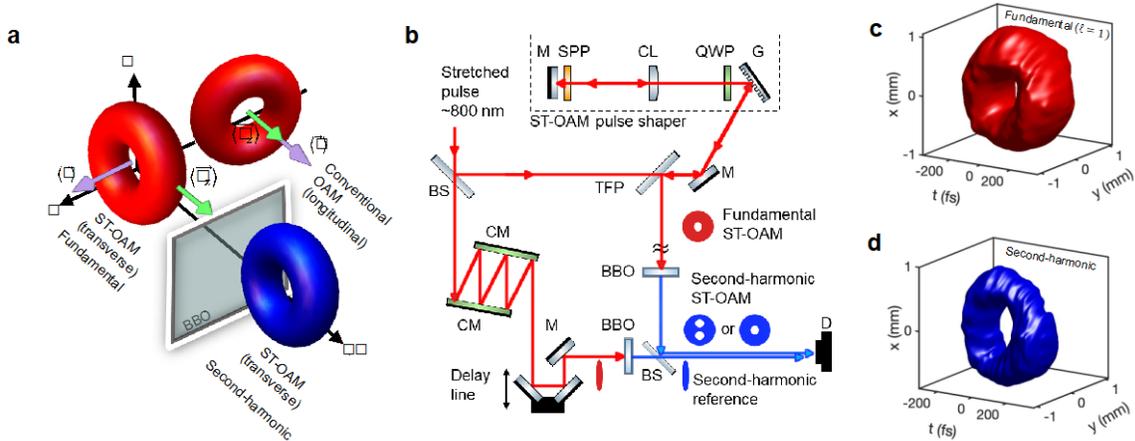

**Fig.1 | Second-harmonic ST-OAM pulse generation and characterization. a.** Schematic of conventional OAM of light, ST-OAM of light, and its SHG. The averaged OAM $\langle\vec{L}\rangle$ (purple arrows) in a conventional OAM pulse is parallel to the mean wavevector $\langle\vec{k_z}\rangle$ (green arrows), and thus named longitudinal OAM, while the averaged OAM in an ST-OAM pulse is perpendicular to $\langle\vec{k_z}\rangle$ and thus transverse. **b.** In our experiment, fundamental ST-OAM pulses of topological charge $\ell = 1$ are generated by a custom pulse shaper. Second-harmonic ST-OAM pulses are generated in BBO crystals and characterized by interference with reference pulses. **c.** Experimentally reconstructed 3D intensity isosurface profile of the fundamental ST-OAM pulse and its second-harmonic shown in **d**. M: mirror; DM: dichroic mirror; CM: chirped mirror; CL: cylindrical lens; BS: beam splitter; TFP: thin film polarizer; QWP: quarter wave plate; SPP: spiral phase plate; G: grating; BBO: beta barium borate crystal; D: detector.

## Results

**The generation and measurement of light with ST-OAM and its second-harmonic.** The generation of ST-OAM pulsed beams is depicted in Fig. 1b. A fundamental pulse at central wavelength $\lambda$ = 800 nm from a Ti:Sapphire amplifier is sent to a custom ST-OAM pulse shaper to generate light with ST-OAM of $\ell$ = 1 (see **Methods**). Unlike conventional OAM beams which can be characterized by space-based methods such as fork holograms[18], coherent diffractive imaging[19], structured apertures[20], ST-OAM pulses require a space-time or equivalently space-frequency based characterization method. Figures 1c and 1d show experimentally measured and reconstructed 3D intensity isosurface profiles of the fundamental and second-harmonic ST-OAM pulse, respectively.

We used a Mach-Zehnder-like scanning interferometer to optically gate an ST-OAM pulse to fully characterize ST-OAM of light, a method similar to that in Ref.[12] Briefly, a long, 800-nm fundamental, ST-OAM pulse of ~500 fs of $\ell$ = 1 was interfered with a short, 800-nm fundamental, Gaussian reference pulse of ~45 fs to form interference fringes. These fringes characterize the portion of the ST-OAM pulse gated by the >10x shorter reference pulse. By scanning the time delay, the spatiotemporal profile of the ST-OAM pulse can be reconstructed computationally from the delay-dependent fringes, where each delay is a time frame of the reconstruction (See **Methods**). Representative fringe patterns of ST-OAM pulses are shown in Supplementary Information (Fig. S2). The reconstructed spatiotemporal amplitude and phase of the fundamental ST-OAM pulse are shown in Fig. 2a-2c. Figure 2a is the amplitude envelope, 2b is the phase, and 2c is a complex field representation where the amplitude and phase are represented by brightness and hue, respectively. Our experimentally reconstructed spatiotemporal profiles clearly confirm the generation of a vortex-shaped ST-OAM pulse with a 2π "azimuthal" phase swirl in space and time.

Second-harmonic fields were generated by sending the fundamental pulses through individual BBO crystals on two arms of a similar interferometer. Therefore, we obtained a long, 400-nm second-harmonic, ST-OAM pulse of an unknown, to-be-determined spatiotemporal topological charge $\ell$, and a short, 400-nm second-harmonic, Gaussian reference pulse to form interference fringes. A second-harmonic ST-OAM pulse generated by a thin BBO crystal (20 μm in thickness) is experimentally reconstructed and shown in Figs. 2d-2f. The fringe patterns used for reconstruction are shown in Supplementary Information (Fig. S2). The amplitude envelope profile of the second-harmonic pulse in Fig. 2d still looks close to a simple vortex shape, which shares the same topology with the fundamental ST-OAM pulse: a torus of genus close to one. Although the reconstructed phase profile in Fig. 2e shows two singularities, they are almost overlapped in the time domain and slightly dislocated in space by only ~200 μm, approximating a single phase singularity with 4π azimuthal spiral phase in space-time. The small singularity separation could be attributed to a slight offset of the laser central wavelength to the design SPP wavelength, and/or a slight amount of astigmatism of the initial Gaussian beam. Also, the generation of the fundamental "spatial" OAM beams inside the pulse shaper is far from perfect—they are generated slightly off the Fourier plane of the 4-f pulse shaper because we relied on the beam double passing through the same SPP by retroreflection (See **Methods**).

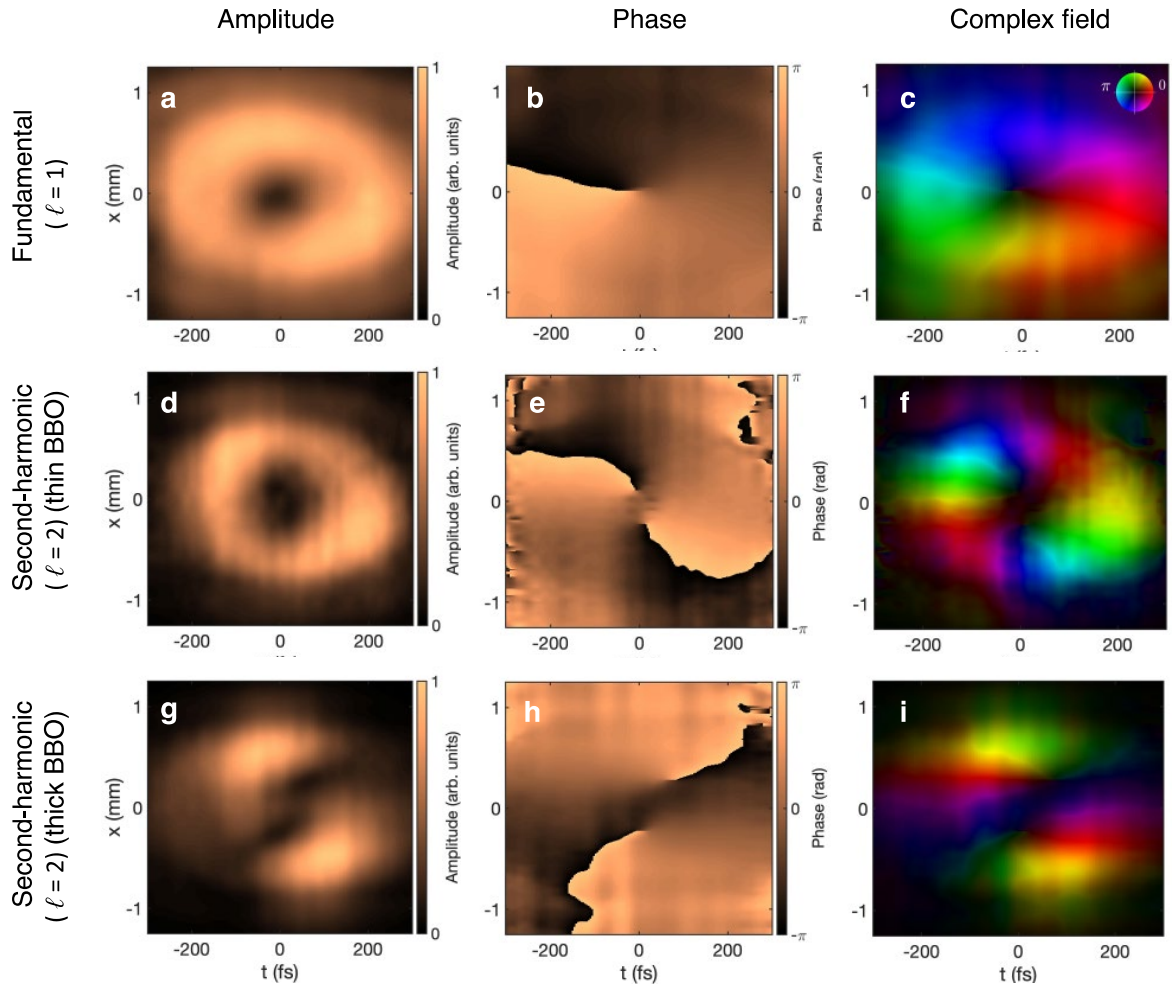

**Fig.2 | Experimentally reconstructed amplitude and phase of ST-OAM pulses.** A fundamental ST-OAM pulse of topological charge ℓ =1, its second-harmonic fields generated by a thin and a thick BBO crystal, are shown in the top, middle, and bottom rows, respectively. Experimentally measured and reconstructed amplitude envelope, phase, and complex field representation are shown in the left, middle, and right columns, respectively. In the complex field representation, the amplitude is represented by the brightness while the phase is represented by the hue of the color wheel. The fundamental ST-OAM pulse shows 2π spiral phase accumulation on traversing a closed spatiotemporal path around the singularity. The second-harmonic fields in both thin and thick BBO crystals show that ℓ is doubled after frequency doubling, evidenced by the 4π phase accumulation. The second-harmonic fields with ℓ =2 demonstrate the conservation of ST-OAM in an SHG process.

The 4π spiral phase we observed indicates that the second-harmonic ST-OAM pulse has a confirmed spatiotemporal topological charge of ℓ = 2 after frequency doubling. Therefore, we established a nonlinear scaling rule and uncovered that the spatiotemporal angular momentum is conserved during an SHG process.

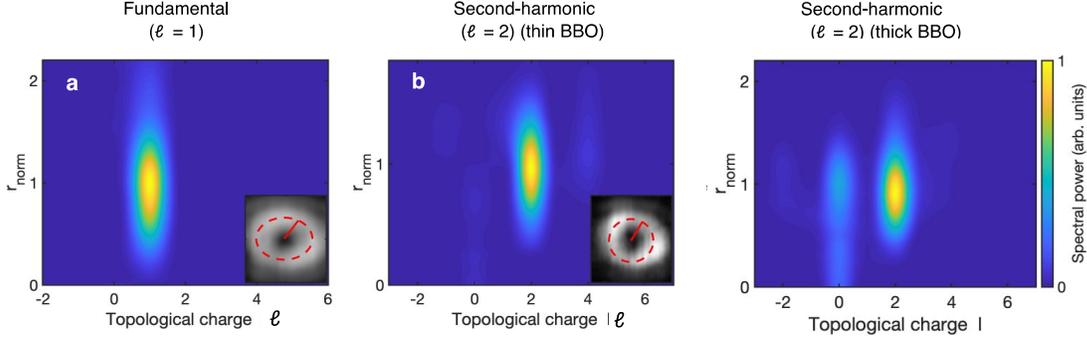

**Fig. 3 | Modal decomposition of the ST-OAM pulses from experiments.** OAM mode spectra of the ST-OAM pulses from **a,** fundamental field, **b,** second-harmonic fields by a thin BBO, and **c,** from a thick BBO crystal. The modal decompositions are obtained by Fourier transformation of the measured spatiotemporal phase along the azimuthal direction on the $(x, t)$ plane for different normalized polar radii $r_{norm}$. The radii are shown in solid red lines in the insets overlapped with their corresponding amplitude envelope profiles for reference. The color map corresponds to the power spectrum in arbitrary units. In **a**, the ℓ = 1 charge contains ~94.5% of the total spectral power. In **b** and **c**, the ℓ = 2 charge contains ~80.9% and ~57.8% of the total power, respectively.

**Spatiotemporal astigmatism and modal analysis.** We also investigated a case with strong spatiotemporal astigmatism by generating second-harmonic ST-OAM pulses using a thick (1 mm) BBO crystal, as shown in Figs. 2g-2i. Unlike the thin BBO case, the second-harmonic pulse from a thick BBO crystal has a different topology in the pulse profile: there are two amplitude holes (two phase singularities) close to the center with two outer lobes — a case resembling a torus of genus two. The phase profile in Fig. 2h shows two phase singularities with a $2\pi$ spiral phase individually, giving rise to a $4\pi$ accumulated phase. This again demonstrates ST-OAM conservation in an SHG process, although the topology is not conserved. The two phase singularities are dislocated in both time, by ~50 fs, and in space, by ~500 μm. We note that this separation is not surprising – a similar deformation was described in an earlier simulation[9], and such separation of phase singularities in an SHG process was predicted and presented in a recent simulation[21]. The use of a thicker BBO crystal might lead to spatiotemporal astigmatism by introducing phase mismatch and group velocity mismatch. The phase mismatch can narrow the frequency conversion bandwidth and the group velocity mismatch can stretch the second-harmonic pulse duration. Both these effects would contribute to the final distortion of the second-harmonic ST-OAM pulses in space and time, causing a change in topology of the vortices. Detailed investigations of such complex processes in thick nonlinear crystals are beyond the scope of this work. Further studies are needed to explore the propagation effects of ST-OAM pulses in nonlinear dispersive media.

To quantify how these ST-OAM pulses are distorted by imperfect SHG conditions, we performed a modal decomposition of the ST-OAM pulses. The OAM mode spectra are extracted by Fourier transformation over the experimentally reconstructed phase profiles along the azimuthal direction on the $(x, t)$ plane for normalized polar coordinates $r_{norm}$. Unlike a conventional OAM beam where the x- and y-axes are equivalent, ST-OAM pulses need to be scaled in the x-axis and the t- (z-) axes to obtain the mode spectrum correctly (See **Methods)**. The insets in Fig. 3 show how the origins in the polar coordinates and scaling factors are defined in each case to deal with a different number of phase singularities, where the solid red lines are overlapped with amplitude envelope profiles taken from Figs. 2a-2i for reference. The mode spectra are shown in Figs. 3a for the fundamental pulses, 3b for the second-harmonic pulses generated using a thin BBO crystal, and 3c for the second-harmonic pulses generated using a thick BBO crystal. As expected, the peak of the ST-OAM spectrum is located at $\ell = 1$ for the fundamental field and mainly at $\ell = 2$ for its second-harmonic field.

To estimate how much spectral power is located at the expected charge number, each mode spectrum was integrated along the radial coordinate. For the fundamental field, ~94.5% of the total spectral power is located at $\ell = 1$ (Fig. 3a). For the second-harmonic field generated using a thin BBO crystal in Fig. 3b, the majority (~80.9%) of the total power is at $\ell = 2$ (Fig. 3b), reflecting the conservation of ST-OAM in the SHG process. However, in the case of a thick BBO crystal, we observed severe degradation or increased impurity of the modal spectrum, where only ~57.8% of the total power is at $\ell = 2$ and a significant DC peak appears (Fig. 3c). This result indicates that the second-harmonic ST-OAM field is distorted after the generation and propagation through the thicker crystal due to spatiotemporal astigmatism.

**The momentum density and energy density flux.** We next investigated the experimentally extracted momentum density and energy density flux of the fundamental and its second-harmonic ST-OAM pulses to better understand the SHG process including spatiotemporal astigmatism. Our ST-OAM pulses, as quasi-monochromatic, near paraxial fields with linear polarization, can be described by a scalar field $E = E_0 e^{i\Phi}$ (see Supplementary Information for details). The *canonical momentum density in vacuum* $\vec{P}$ (also called scalar optical current in free-sapce[22], with the dimension of energy per unit time and per unit area) is then defined as the expectation value of the intensity-weighted momentum operator: $\vec{P} \propto \frac{1}{\omega} \text{Im}(E^* \nabla E) = \frac{1}{\omega} |E_o|^2 \nabla \Phi$.[3,22,23] It was shown that the canonical momentum density can describe the local phase gradient in an arbitrary structured field and is much more suitable to describe momentum and angular momentum properties of free-space light field than the kinetic Poynting vector[3,24]. Such momentum density can be further used to extract a single quantity, the normalized expectation value of the ST-OAM per pulse $\langle \ell \rangle = \omega |\int \vec{r} \times \vec{P} \, dV|/(\int E^* E dV)$, following Ref.[12] Adopting these definitions and assuming unity refractive index in air, Figs. 4a-4c show our experimentally measured spatiotemporal canonical momentum density for a fundamental pulse in 4a, and its second-harmonic pulse generated using a thin and a thick BBO crystal in 4c and 4b. respectively. The modulus of the momentum density vectors ($|\vec{P}|$, the length of the yellow arrows in the figures) are scaled by $1/ct_o$ and by $1/x_0$ in the horizontal and vertical axes for a better visualization (see Methods). The momentum density of a fundamental ST-OAM pulse follows the gradient of the spatiotemporal phase and form a vortex around the singularity. For a second-harmonic pulse in a thin BBO crystal, the momentum density is no longer uniformly distributed, but it still forms a vortex. In this case, we obtain the ratio of the normalized expectation value of ST-OAM per pulse $\langle \ell^{(2\omega)} \rangle / \langle \ell^{(\omega)} \rangle = 2.09$, the scenario of a near perfect SHG process. However, for the case in a thick BBO crystal, the momentum density is redistributed following the distorted amplitude profile, resulting in the reduced ratio $\langle \ell^{(2\omega)} \rangle / \langle \ell^{(\omega)} \rangle = 1.36$. The smaller expectation value of ST-OAM per pulse is consistent with the appearance of a significant DC peak in the OAM modal spectrum observed in Fig. 3c, similarly reflecting the degradation and the reduced mode purity due to spatiotemporal astigmatism in a thick crystal.

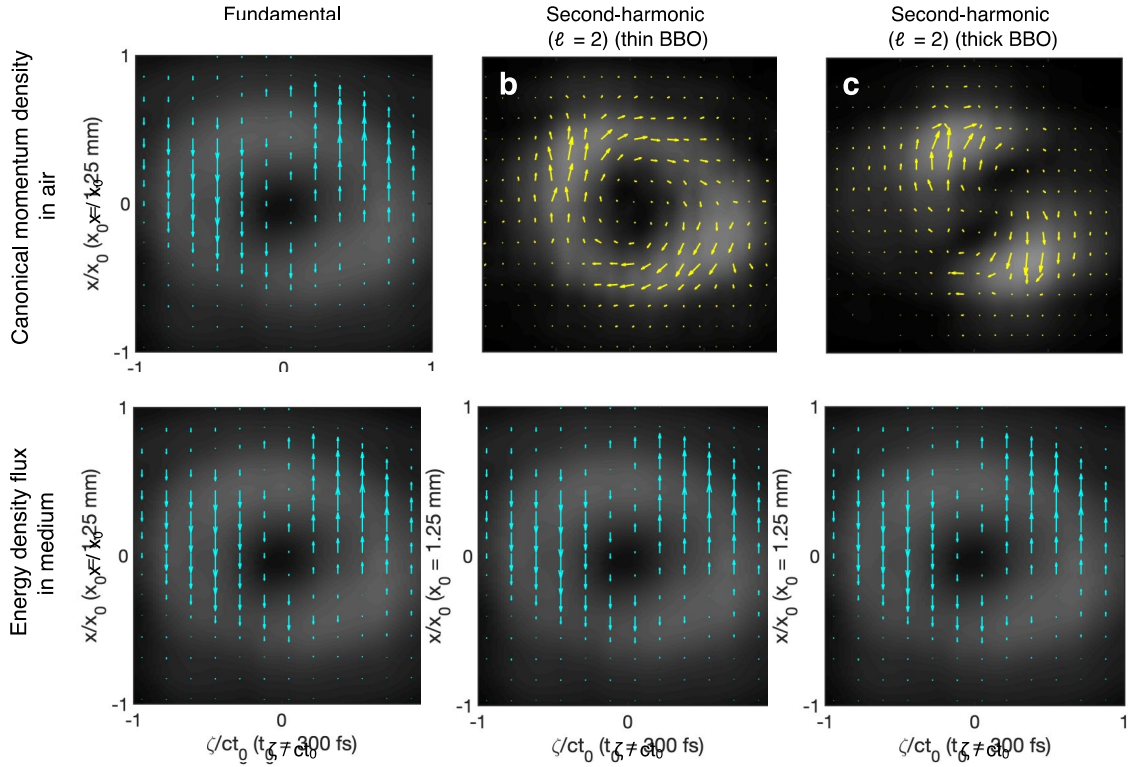

**Fig.4 | Experimentally extracted spatiotemporal momentum density and energy density flux of ST-OAM pulses.** A canonical momentum density in air as defined in Ref[3,22,23] is shown in **a,** a fundamental ST-OAM field, **b,** its second-harmonic field generated by a thin BBO, and **c,** by a thick BBO crystal. An energy density flux in dispersive media (BBO crystals), which is proportional to both phase gradient and material dispersion as defined in Ref,[25,26] is shown in **d,** a fundamental ST-OAM field, **e,** its second-harmonic field generated by a thin BBO, and **f,** by a thick BBO crystal. The modulus of the canonical momentum density (the length of the yellow arrows) is scaled by $1/ct_o$ and $1/x_0$ along the horizontal and vertical axis, respectively, for better visualization. Contrarily, the modulus of the energy density flux (the length of the cyan arrows) is not scaled. The ST-OAM pulses propagate from left to right.

To further understand the physical mechanism responsible for this astigmatism, we experimentally extracted the *energy density flux* in the ST-OAM pulses *inside* the BBO crystals. We note that energy flows in electromagnetic waves are generally described by the well-known Poynting vector, which coincides with the canonical momentum density in air or free-space under scalar monochromatic field approximation. It is worth mentioning that there have been long-lasting debates on the correct expression of energy, momentum, and angular momentum properties in dispersive media[24,27,28]. Thus, here we used another convenient, measurement-friendly definition called *energy density flux in medium,* as is used and introduced in Ref,[25,26] to characterize the local distributions of optical energy per unit time and per unit area inside the dispersive BBO crystals. This energy density flux $\vec{J}$ separates the transverse $(x, y)$ and longitudinal $(\zeta; z, t)$ contributions and can be written as $\vec{J} \propto \frac{1}{\omega}|E_o|^2[\nabla_\perp \Phi - \text{GVD}(\partial\Phi/\partial\zeta)\hat{\zeta}]$, following Ref,[10,26] where $\nabla_\perp$ is the transverse gradient and GVD is the dimensionless group velocity dispersion of the material (see **Methods**). In Figs. 4d-4f, the modulus of the energy density flux ($|\vec{J}|$, the length of the cyan arrows in the figures) is presented without rescaling, so we can observe its orientation directly. In a fundamental ST-OAM pulse, the energy density flux forms a saddle-shaped antivortex structure around the singularity, given the positive GVD at this wavelength of a BBO crystal. As a fundamental ST-OAM pulse propagates through a thin BBO crystal, the SHG process, as well as a small amount of dispersion, leads to minor energy redistribution. In contrast, in the case of a thick BBO crystal, larger dispersion can significantly reshape the ST-OAM pulses, leading to strong spatiotemporal astigmatism in the second-harmonic ST-OAM pulse. In Figs. 4d-4f, our extracted energy density flux in the BBO crystal shows a tendency to reshape the ST-OAM pulse into a two-lobe structure upon continued propagation, resulting in the higher degree of distortion observed for the pulse emerging from the thick BBO crystal. It is worth mentioning that the canonical momentum density in air (Figs. 4a-4c) represents the nontrivial optical power distributions of the fields generated *after* the nonlinear conversion, while the energy density flux in media (Figs. 4d-4f) represents the nontrivial optical power distributions of the fields *inside* the dispersive BBO crystals.

## Discussion

The second-harmonic ST-OAM pulse generated using a thin BBO crystal can be described by a simple theory. In this simple scenario with perfect phase matching, an undepleted pump approximation, and assuming a lossless medium, the second-harmonic field is proportional to the square of the complex input fundamental field, namely,

$$E^{(2\omega)}(x,y,\zeta) = E_0 e^{i\ell^{(2\omega)}\Phi(x,y,\zeta)} \propto [E^{(\omega)}(x,y,\zeta)]^2 \propto [e^{i\ell^{(\omega)}}]^2 \propto e^{i2\ell^{(\omega)}}. \tag{2}$$

Here $\zeta \equiv z - \mathrm{v}_g t$ is the space coordinate in the moving reference frame of a pulse traveling with the group velocity $\mathrm{v}_g$ and the definition of the scalar field follows Equation (1). As a result, $e^{\ell^{(2\omega)}} = e^{2\ell^{(\omega)}}$ and we obtain $\ell^{(2\omega)} = 2\ell^{(\omega)}$. This indicates that the space-time topological charge of the fundamental field is doubled along with the optical frequency. In this scenario, the field profile of the second-harmonic wave should mimic the fundamental pulse. Because the pulse duration and the beam size of a fundamental pulse can be shortened and reduced in a perfect SHG process due to its quadratic response, it partially explains the fact that the second-harmonic ST-OAM pulse appears thinner when plotting intensity isosurface profiles as shown in Figs. 1 and Figs. 2. The nonlinear relationship between the electric field amplitudes of a fundamental and its second-harmonic field means that the amplitude profile of the ST-OAM pulse changes during frequency doubling. Beyond the simple scenario, the SHG process is further complicated by additional factors in addition to phase mismatches, such as group velocity mismatch, intrapulse group velocity dispersion, absorption or losses, and pump depletion[29,30]. A rigorous theoretical study of a final second-harmonic ST-OAM pulse profile must take the above factors into account and is the subject of further detailed investigation, which is beyond the scope of this manuscript.

Importantly, our demonstration of the ST-OAM nonlinear conversion and the ST-OAM conservation in SHG shows that a new avenue of secondary sources with ST-OAM can be realized experimentally in up-converted and likely down-converted electromagnetic waves. For example, it would be of interest to investigate the stimulated OAM Raman scattering,[31] harmonic spin-orbit angular momentum cascade,[32] or entangled photons generated by parametric processes with ST-OAM pulses. It is also possible that other matter waves such as electron vortex beams or electron bunches with ST-OAM can be generated through photoelectron generation (photoemission) or other means that could improve accelerator technologies and electron microscopy[33] in the future.

In summary, we report the experimental observation of a second-harmonic ST-OAM of light and its space-time topological charge conversation during frequency doubling from $\ell = 1$ to $\ell = 2$. The charge of the ST-OAM pulse, corresponding to the space-time spiral phase structure, was observed to double in an SHG process. Our finding also confirms that the spatiotemporal phase singularities in ST-OAM of light can be interpreted as space-time topological charges carrying transverse OAM— the term coined in analogy to the spatial topological charges in a conventional OAM of light carrying longitudinal OAM. In this analogy, both types of OAM follow the same charge conservation and nonlinear scaling rules. We also found that the topological structure of the space-time phase swirl in a second-harmonic field may not be conserved. An SHG process can generate additional phase singularities, depending on spatiotemporal astigmatism due to group velocity mismatch or phase mismatch condition in a dispersive medium. SHG is the foundation of any nonlinear optics textbook. Our work thus opens a new avenue of light carrying ST-OAM in nonlinear conversion and scaling, and it further suggests the possibilities to drive secondary ST-OAM sources from electromagnetic waves, mechanical or acoustic waves, to matter waves.

*Footnote Added*: It has come to our attention that *after* our manuscript was submitted and the preprint was posted online in an open repository (Gui *et al*., Second-harmonic generation and the conservation of spatiotemporal orbital angular momentum of light, 10 December 2020, PREPRINT (Version 1) available at Research Square [https://doi.org/10.21203/rs.3.rs-116263/v1]), an independent experiment that explored similar light science using a different experimental approach was reported in a preprint online (arXiv:2012.10806v1).

## Acknowledgments

The authors gratefully acknowledge funding from an AFOSR MURI (FA9550-16-1-0121). N. J. B. acknowledges support from National Science Foundation Graduate Research Fellowships (Grant No. DGE-1650115).

## Author Contribution Statement

C.-T.L. conceived the project. G.G. conducted and designed the experiment. C.-T.L. and G.G. both analyzed the data. M. M. M. and H. C. K. proposed the research thrust, supervised the research, developed the generation and measurement capabilities, and applications. All authors contributed to the discussion and writing of the manuscript.

## Competing Interest Statement

M. M. M. and H. C. K. have a financial interest in KMLabs. The other authors declare no competing financial interests.

## Methods

**The ST-OAM pulse generation.** A stretched (full width at half maximum pulse duration ~500 fs) optical pulse at a central wavelength ~800 nm from a regenerative Ti:sapphire amplifier at 1 kHz repetition rate (KMLabs Wyvern HE) was first split into two paths before entering into a homemade Mach-Zehnder-type scanning interferometer. One path is for generating a short (~45 fs), Gaussian reference pulse, and the other is for generating a long (~500 fs) ST-OAM pulse. The reference Gaussian pulses were compressed using dispersive chirped mirrors (Ultrafast Innovations) with multiple bounces to shorten the pulse duration. The ST-OAM pulses were generated by a custom pulse shaper we designed, consisting of a reflective grating (600 groove/mm), a cylindrical lens (f = 25 cm), a multi-faceted spiral phase plate (SPP, HoloOr, 16-steps per phase wrap), and a high-reflectance end mirror for retroreflection. To generate fundamental ST-OAM pulses of the *spatiotemporal* topological charge $\ell = 1$, an SPP with a designed *spatial* topological charge $\ell_c = 0.5$ at the design wavelength 790 nm was used. Upon a retroreflection from the end mirror, the pulse passes through the SPP twice and generates the desired spatial topological charge $\ell_c = 0.5 - (-0.5) = 1$. Note that an SPP is a non-reciprocal optical component, and a mirror reflection changes the sign of the $\ell_c$. Thus, the double-passing of the SPP from the opposite direction makes the $\ell_c$ charge addition possible, when the two passes are very close. A Fourier transform in the spectral domain of the pulse shaper converts a spatially chirped beam with a spatial spiral phase into a pulsed beam with a spatiotemporal spiral phase. BBO crystals with thickness 20 μm (thin BBO) and 1 mm (thick BBO), cut and oriented to phase match type-I SHG, are used to generate the second-harmonic fields at wavelength ~400 nm, which are placed in the far field from the pulse shaper. The second-harmonic Gaussian reference pulse was generated by another 200-μm thick type-I SHG BBO crystal. To avoid any distortion and diffraction of the spatiotemporal phase structure after focusing the pulse, we used a nearly collimated ST-OAM beam in the SHG process, without additional optics that will cause astigmatism in space and time[34].

**The amplitude and phase reconstruction.** Both fundamental and second-harmonic field amplitude $|E_0(x, y, t)|$ and phase $\Phi(x, y, t)$ were measured using a Mach-Zehnder-type scanning interferometer. When scanning the time delay between a short reference pulse and a long ST-OAM pulse, the reference pulse serves as an optical gating pulse that provides spatial information of both amplitude and phase at the given time delay. We used 38 delays to reconstruct fundamental ST-OAM pulses and >50 delays for second-harmonic ST-OAM pulses. Representative fringe patterns are shown in Supplementary Information (Fig. S2). To reconstruct the 3D amplitude profile $|E_0(x, y, t)|$, we first extracted the envelope of the fringe patterns at various time delays. At each time delay, the fringe envelope was then divided by the amplitude of the reference beam, which gives the 2D amplitude of the ST-OAM pulse at each delay. By stacking 2D amplitude profiles at various delays, a 3D amplitude reconstruction is retrieved. We then used them to reconstruct and present 3D intensity isosurface profiles, with 37% energy of the pulse contained in the torus-shaped isosurface, as shown in Fig. 1c and 1d. On the other hand, the 2D amplitude profiles shown in Fig. 2 are projections along the y-axis of the 3D amplitude profiles on to the $(x, t)$ plane, $|E_0(x, t)|$. To reconstruct the phase, we first applied 1D Fourier transform to the original fringe patterns along the y-axis. By extracting the phase of the AC peak in the Fourier domain, we obtained a 1D phase profile from each 2D fringe pattern, representing $\Phi(x)$ at each delay. The 2D phase reconstruction is retrieved by stacking 1D profiles at each delay. It is worth mentioning that due to the instability of the interferometer, the position of the fringe patterns can vary shot-to-shot. To overcome this problem, the delay-dependent 1D phase profiles were numerically shifted by constant phases, which are calculated by the amplitude-weighted least square method to achieve the best phase continuity (smoothness) in the time domain. Besides, all phase profiles were subtracted by the same background to minimize linear phases and the trivial temporal phase $e^{-i\omega t}$ is omitted. Both amplitude and phase reconstructions in Fig. 2 were further interpolated for better visualization.

**The ST-OAM modal decomposition.** OAM mode spectra are obtained by calculating the azimuthal Fourier transform of the measured spatiotemporal phase along the azimuthal direction on the $(x,t)$ plane using normalized polar coordinates, $r_{norm} = \sqrt{(x/x_w)^2 + (ct/ct_w)^2}$ and $\phi_{norm} = \tan^{-1}[(x/x_w)/(ct/ct_w)]$. To calculate the normalization factors $x_w$ and $t_w$, we first defined the origin of the polar coordinate as the position of the phase singularity (or the midpoint of multiple singularity positions), corresponding to the central minimum in the amplitude profile. Then a cross-section of the vortex, including the origin, was plotted along the x-axis and the t-axis. These cross-sections show two peaks, i.e., the front and rear ends of the pulse. We then define $x_w$ and $t_w$ to be one-half of the peak-to-peak separation along the x- and the t-axis, respectively. In Fig. 3c, the scaling factors are inherited from those of Fig. 3b, since the amplitude profile shown in Fig. 3c is not in an ideal vortex shape. The spectral resolution of the mode spectrum is limited to unity, a quantized integer, the averaged expectation value of the spatiotemporal topological charges. The spectrum plots are interpolated for smoothness.

**Experimentally extraction of the momentum density and energy density flux.** The free-space *canonical momentum density in vacuum* or in air, also called scalar optical current or wave current, is defined as the expectation value of the intensity-weighted momentum operator: $\vec{P}(x,y,z,t) \propto \frac{\varepsilon}{\omega} \text{Im}(E^* \nabla E) = \frac{\varepsilon}{\omega} |E_o(x,\zeta)|^2 \nabla \Phi(x,\zeta)$, where $\varepsilon$ is permittivity. This definition is valid under paraxial-wave approximation (collimated beam with divergence angle <<1 radian; our beam divergence is <0.001 radians), scalar field approximation (linearly polarized light), quasi-monochromatic field approximations (relative bandwidth Δω/ω=Δλ/λ=30nm/800nm=0.0375<<1), and with negligible magnetic field contribution. The quasi-monochromaticity assumes that the field envelopes vary slowly on the time scale of an optical period, which is also the case in our experiment where the optical period divided by the field envelope is ~2.7fs/300fs=0.009<<1. Using this definition and approximations, we can then calculate the normalized expectation value (not a quantum number) of the ST-OAM per pulse $\langle \ell \rangle = \omega \left| \int \vec{r} \times \vec{P} \, dV \right| / (\int E^* E dV)$, where the "volume" integral is $\int dV = \int dx d\zeta$. In the definition of $\langle \ell \rangle$, the numerator term is the expectation value of angular momentum, and the denominator term is the expectation value of the energy. For a polychromatic field, the canonical momentum density and the associated expectation value of the OAM are ill-defined. In our analysis, the gradient operation is achieved by numerically calculating the differential of the phase and applying a moving average to the resultant vector field to remove high frequency noises. The modulus of the canonical momentum density vector $|\vec{P}|$, the length of the yellow arrows in Fig. 4a-4c, are scaled by $1/x_0$, $x_0$=1.25 mm, on the x-axis, and by $1/ct_0$, $t_0$=300 fs, on the $\zeta$-axis for better visualization.

The *energy density flux in dispersive media* was used and introduced in Ref[25,26] to characterize the propagation of ultrashort laser pulses with spatiotemporal coupling. It has been used over the past decade in applications such as filaments and light bullets propagating in dispersive media. It is defined as $\vec{J}(x,y,z,t) \propto \frac{1}{\omega}|E_o(x,y,\zeta)|^2[\nabla_\perp \Phi(x,y) - \text{GVD}(\partial \Phi(\zeta)/\partial\zeta)\hat{\zeta}]$ with transverse and longitudinal terms separated as in Ref,[10,26] where $\nabla_\perp$ is the transverse gradient and GVD is the dimensionless group velocity dispersion of the material. The modulus of the energy density flux vector $|\vec{J}|$, the length of the cyan arrows in Fig. 4d-4f, are not scaled to show the original orientation of the flux directly. Note that the canonical momentum density is experimentally extracted based on the measured spatiotemporal phase on the detector plane in air. On the other hand, the energy density flux is experimentally extracted based on the same measured spatiotemporal phase on the detector plane in air, while taking the dispersive medium property into account. In both cases, we ignored the short distance propagation from the BBO crystals to the detector plane that leads to a small and negligible propagation-induced quadratic phase. Also, in both cases, the trivial linear momentum density and energy density flux in the t- and z-axis are removed, to better illustrate the important optical power distributions in space-time in three different scenarios, comparing the fundamental field, the second-harmonic field from a thin BBO crystal, and the second-harmonic field from a thick BBO crystal.

# Supplementary Information

## 1. FROG measurement of fundamental ST-OAM pulses

     A common SHG frequency-resolved optical gating (FROG) setup (videoFROG, MesaPhotonics) can also be used to measure the spatiotemporal phase structure of an ST-OAM pulse (Fig. S1**a**) by carefully placing an entrance pinhole at different positions of the beam and scanning over it. Figure S1**b** and S1**c** show a typical FROG trace and its reconstruction from a fundamental ST-OAM pulse of $\ell = 1$ at central wavelength $\lambda = 800$ nm and measured at the beam center (Fig. S1**a**). In Fig. 1**c**, the retrieved field amplitude envelope $|E_0(x = 0, y = 0, t)|$ clearly shows a double-pulse-like structure in the time domain. Also, the retrieved temporal phase $\Phi(x = 0, y = 0, t)$ shows a clear $\pi$ phase jump between the front and the rear ends, consistent with the expected $2\pi$ phase shift along the azimuth in the $(x, t)$ plane and a centered phase singularity. In the FROG measurements, we used a compressed ST-OAM pulse with a shorter pulse duration for easier implementation in the FROG setup, in comparison with the longer ST-OAM pulse shown in Fig. 1c. In principle, one can scan an entrance pinhole over the beam profile to get a spatially resolved FROG measurement. Although doable, this method is tedious with slow throughput. In addition, SHG-FROG is known to have ambiguities on nontrivial time direction, absolute phase, and temporal shift. Therefore, we used a Mach-Zehnder-like scanning interferometer to optically gate an ST-OAM pulse to fully characterize an ST-OAM of light and its second-harmonic as shown in the Main Text.

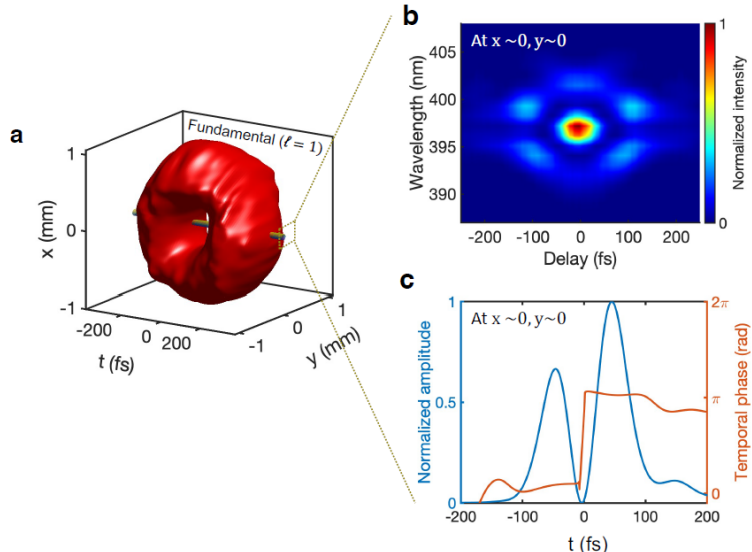

**Fig. S1. FROG measurement of ST-OAM pulses. a.** Experimentally reconstructed 3D intensity isosurface profile of an 800 nm fundamental ST-OAM pulse using scanning interferometry**.** The center part of the fundamental ST-OAM pulse, indicated by the brown bar, was experimentally sampled and charactered by an SHG-FROG setup, where the measured FROG trace and the retrieved temporal field envelope and the temporal phase are shown in **b** and **c**, respectively.

## 2. Experimental fringe patterns measured by scanning interferometry

Experimental fringe patterns of fundamental and second-harmonic ST-OAM pulses are recorded by a Mach-Zehnder-type scanning interferometer. From the recorded data, the fringe shift between the upper and lower parts of the pattern indicates a spatial phase shift in the ST-OAM pulse. Some representative fringe patterns of the fundamental ST-OAM pulses are shown in Fig. S2**a**, with two black markers added to each delay time frame as a visual guide to the eye to observe the fringe shift. The discontinuous phase jump (or phase dislocation) around the singularity happens close to delay zero. When the reference pulse was scanned towards the ST-OAM phase singularity from -300fs to 0fs, the fringes are first tilted, and then shifted by a $\pi$ phase difference at the center of the singularity. When the reference pulse was scanned away from 0fs to 300 fs, the fringes become straight again, but the upper and lower fringes are shifted by one fringe. Such a fringe shift is a feature of an ST-OAM pulse of $\ell = 1$. Figure S2**b** shows the fringe patterns of second-harmonic ST-OAM pulses generated by a thin BBO crystal (20 μm in thickness), where we can see the upper and the lower parts of the pattern are shifted by two fringes, indicating a $4\pi$ phase swirl around the singularity. Similarly, Fig. S2**c** shows the fringe patterns of second-harmonic ST-OAM pulses generated by a thick BBO crystal (1 mm in thickness), where the upper, middle, and lower parts of the pattern are shifted at various delay time and spatial locations, indicating the separation of phase singularities.

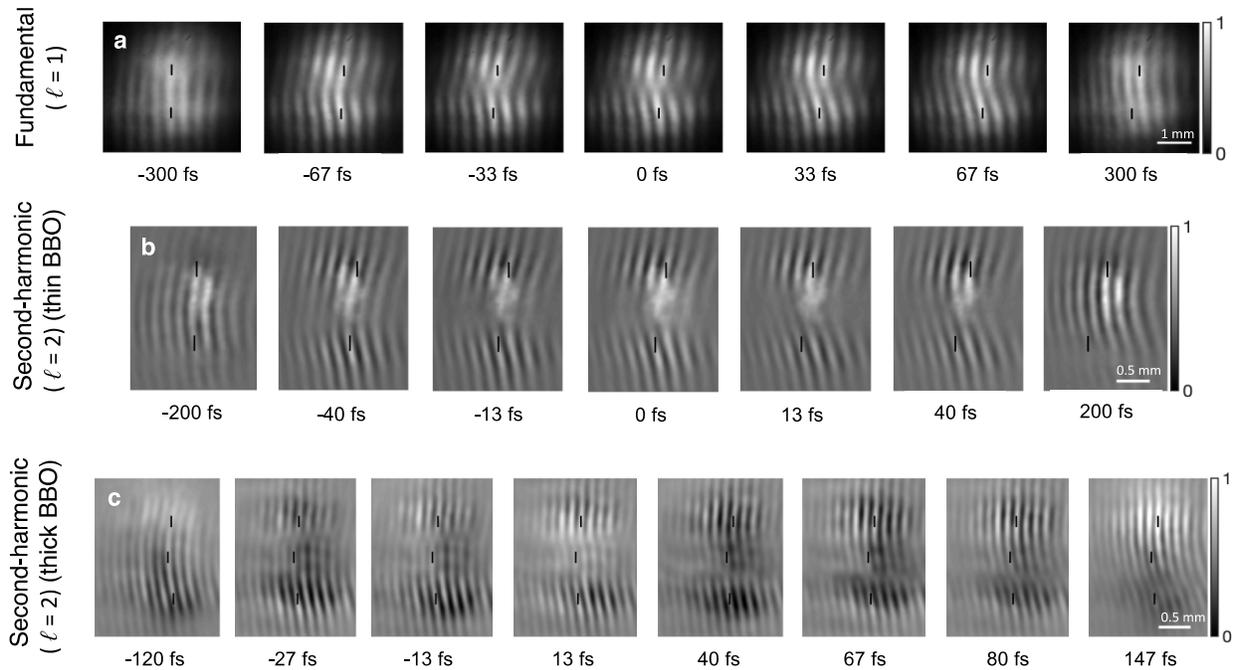

**Fig. S2. Measured fringe patterns of ST-OAM pulses by scanning interferometry. a.** Representative fringe patterns of fundamental ST-OAM pulses at various time delays, with two black markers added as an aid to the eye to observe the fringe shift. **b**. Representative fringe patterns of second-harmonic ST-OAM pulses generated by a thin BBO crystal. The result that the fringes are shifted by two shows that the spatiotemporal topological charge is doubled in an SHG process. **c**. Similar to the case in **b** but with a thick BBO crystal. The upper, middle, and lower parts of the pattern are shifted by one fringe from each other at various delay time and spatial locations.